\documentstyle[preprint,aps,12pt,epsfig,floats]{revtex}
\tightenlines

\begin{document}
\newcommand{\beq}{\begin{equation}}
\newcommand{\eeq}{\end{equation}}
\newcommand{\beqa}{\begin{eqnarray}}
\newcommand{\eeqa}{\end{eqnarray}}
\newcommand{\fr}{\frac}
\newcommand{\sect}[1]{\section{#1}\setcounter{equation}{0}}
\renewcommand{\theequation}{\arabic{section}.\arabic{equation}}
\newcommand{\rf}[1]{\ref{#1}}
\preprint{INJE-TP-05-05, hep-th/0506096}

\title{No Hawking-Page phase transition in three dimensions}

\author{ Y. S. Myung\footnote{Email-address :
ysmyung@physics.inje.ac.kr}}
\address{Relativity Research Center, Inje University, Gimhae 621-749, Korea \\
and  Institute of Theoretical Science, University of Oregon,
Eugene, OR 97403-5203, USA}

\maketitle
\vspace{5mm}

\begin{abstract}
We investigate whether or not the Hawking-Page phase transition is
possible to occur in three dimensions. Starting with the simplest
class of Lanczos-Lovelock  action, thermodynamic behavior of all
AdS-type black holes  without charge falls into two classes:
Schwarzschild-AdS black holes in even dimensions and Chern-Simons
black holes in odd dimensions. The former class  can provide the
Hawking-Page transition between Schwarzschild-AdS black holes and
thermal AdS space. On the other hand,  the latter class is
exceptional and thus the Hawking-Page transition is hard to occur.
In three dimensions, a second-order phase transition might occur
between the non-rotating BTZ black hole and the massless BTZ black
hole (thermal AdS space), instead of the first-order Hawking-Page
transition between the non-rotating BTZ black hole and thermal AdS
space.
\end{abstract}

\newpage

\sect{Introduction} Hawking's semiclassical analysis for the black
hole  radiation suggests that most of information in initial
states is shield behind the event horizon and is never back to the
asymptotic region far from the evaporating black hole\cite{HAW1}.
This means that the unitarity is violated by an evaporating black
hole. However, this conclusion has been debated ever
since\cite{THOO,SUS,PAG}. It is closely related  to the
information loss paradox which states the question of whether the
formation and subsequent evaporation
 of a black hole is unitary. One of the most urgent problems in the black
hole physics is to resolve the unitarity issue.

Recently, Maldacena proposed that the unitarity can be restored if
one takes into account the topological diversity of gravitational
instantons with the same AdS  boundary in three-dimensional
gravity\cite{MAL}.  Actually, three-dimensional gravity\cite{BTZ}
is not directly related to the information loss problem because
there is no physically propagating degrees of freedom\cite{CAL}.
If this gravity is part of string theory\cite{HH}, the AdS/CFT
correspondence\cite{MGW} means that the black hole formation and
evaporating process should be unitary because its boundary can be
described by a unitary CFT. On later, Hawking has withdrawn his
argument on information loss and suggested  that the unitarity can
be preserved  by extending Maldacena's proposal to
four-dimensional gravity system\cite{HAW2}. In other words, the
topological diversity is credited with the restoration of S-matrix
unitarity in the formation and evaporation
 of a black hole. In this approach  the thermal AdS space
plays an important role in restoring unitarity.

 We remark an interesting
phenomenon in the AdS black hole thermodynamics. There exists the
Hawking-Page transition between AdS-Schwarzschild black hole and
thermal AdS space in four dimensions\cite{HP}. This transition was
based on the semi-classical approximation of the Euclidean path
integral for the black hole thermodynamics.  Some authors have
proposed  that this transition is possible  in three-dimensional
spacetimes: transition occurs between the non-rotating BTZ black
hole and  thermal AdS space\cite{KPR}. Since the three-dimensional
gravity and its boundary CFT can provide a prototype to compute
bulk thermodynamic quantities and boundary thermal correlators
exactly, they play an important role for investigating the
evaporation of the black hole through the Hawking radiation.

In this letter we  show that it is hard to occur the first-order
Hawking-Page transition between the non-rotating BTZ and thermal
AdS space. Instead, second-order transitions between the
non-rotating BTZ black hole and the massless BTZ black hole
(thermal AdS space) might be  possible to occur in three
dimensions.

For this purpose,  we use the higher-order gravity theory  in any
dimension which may include quantum effects. According to a work
of Ref.\cite{CTZ}, all thermodynamic behaviors of AdS black holes
were classified into two types: AdS-Schwarzschild black holes in
even dimensions and Chern-Simons black holes in odd-dimensions.
The first type possesses a continuous mass spectrum whose vacuum
with zero mass is the thermal  AdS space, whereas the second one
has a continuous mass spectrum whose vacuum is the massless
extremal AdS black hole.

In this work  we consider three interesting
cases\cite{CLZ,CC,MYU,ML}. 1) The non-rotating BTZ black hole
(NBTZ) with $M>0,J=0$: $r_{+}=l\sqrt{M},~T_{H}=\fr{r_+}{2\pi
l^2},~C_{J}=4\pi r_+=S_{BH}$. 2) The thermal AdS  spacetime (TADS)
with $M=-1,J=0$. We
 choose $T_{H}=0,~C_{J}=0,S_{BH}=0$  because of the absence of
the event horizon. This case corresponds to  the spacetime picture
of the NS-NS vacuum state\cite{MS}. 3) The massless BTZ black hole
(MBTZ) with $M=J=0$: $T_{H}=0,~C_{J}=0,S_{BH}=0$. This  is called
the spacetime picture of the RR vacuum state. Although the
thermodynamic properties of TADS and MBTZ are nearly the same,
 their Euclidean topologies are quite different: TADS (MBTZ) are
topologically trivial (non-trivial).

 The organization of this letter is as follows. Section II is devoted to studying  all AdS
black holes in the simplest framework of Lanczos-Lovelock gravity.
 In section III, we show that
there is no Hawking-Page transition  in three dimensions.

\sect{Thermodynamics of AdS black holes} We start with the
simplest class of Lanczos-Lovelock theory in $d$-dimensional AdS
spacetimes\cite{CTZ}

\beq I_k=\kappa \int \sum^k_{p=0} c^k_p {\cal L}^{(p)}
\label{1eq3}
 \eeq
with
 \beq
 \label{2eq3}
 c^k_p=\fr{l^{2(p-k)}}{d-2p}{k \choose p}, ~ p\le k;~~ c^k_p=0,~p>k.
 \eeq
 Here an integer $k$ with $1 \le k\le [(d-1)/2]$ represents the highest
 power of curvature in the lagrangian and
 a parameter $\kappa$  is related to the $d$-dimensional gravitational
 constant $G_k$ as
 $\kappa=1/2(d-2)!\Omega_{d-2}G_k$.
For a given $d$, $c^k_p$ describes a family of inequivalent
theories labelled by $k$.
 ${\cal L}^{(0)}$ is given by the cosmological constant
 $\Lambda=-(d-1)(d-2)/2l^2$ with the AdS curvature radius $l$ and
${\cal L}^{(1)}$ is the curvature scalar  $R$.
 ${\cal L}^{(2)}$ corresponds to the Gauss-Bonnet term of
 $-R_{\mu\nu\alpha\beta}R^{\mu\nu\alpha\beta}+4R_{\mu\nu}R^{\mu\nu}-R^2$,
 whose equation of motion  contains no more than second-order
 derivatives of the metric and which appears to be  a ghost free
 theory. This term is relevant to the case with $d\ge 5$.
Further, $I_1$  corresponds to the $d$-dimensional
Einstein-Hilbert action including the negative cosmological
constant $\Lambda$. If $d=2n-1(n \in {\bf N})$, the maximum value
of $k$ is given by  $k=n-1$. In this case the corresponding
lagrangian is the Chern-Simons $(2n-1)$-form.  For $d=2n$ and
$k=n-1$, the action is defined by the 2-form Pfaffian and it takes
the Born-Infeld form. For  $d=3$ and $d=4$, $I_1^{d=3}$ is the
Einstein-Hilbert action which is equivalent to the Chern-Simons
action and $I_1^{d=4}$ is the Einstein-Hilbert action which
coincides with the Born-Infeld action  upto the Euler density.
 In the case of $d=5$, $I_1^{d=5}$
is the Einstein-Hilbert action, while $I_2^{d=5}$ leads to the
Chern-Simons action.  For $d=6$, $I_1^{d=6}$ is the
Einstein-Hilbert action but $I_2^{d=6}$ is the Born-Infeld action.

 The black hole solution  to the action of Eq.(\ref{1eq3}) is given by\cite{CTZ,ACB}
\beq
 \label{3eq3}
ds^2=
-\Big[1+\fr{r^2}{l^2}-\Big(\fr{2G_kM+\delta_{d-2k,1}}{r^{d-2k-1}}\Big)^{1/k}\Big]dt^2
+\fr{dr^2}{1+\fr{r^2}{l^2}-\Big(\fr{2G_kM+\delta_{d-2k,1}}{r^{d-2k-1}}\Big)^{1/k}}+r^2d\Omega_{d-2}.
 \eeq
The mass of this black hole takes the form \beq \label{4eq3}
M(r_+)= \fr{r_+^{d-2k-1}}{2G_k}\Big(1+\fr{r_+^2}{l^2}\Big)^k
-\fr{\delta_{d-2k,1}}{2G_k} \eeq which is a monotonically
increasing function of the horizon radius $r_+$. The presence of
the Kronecker delta function  shows that there exist two different
vacua ($M=0$) with different causal structures.
\begin{figure}[t!]
\epsfig{file=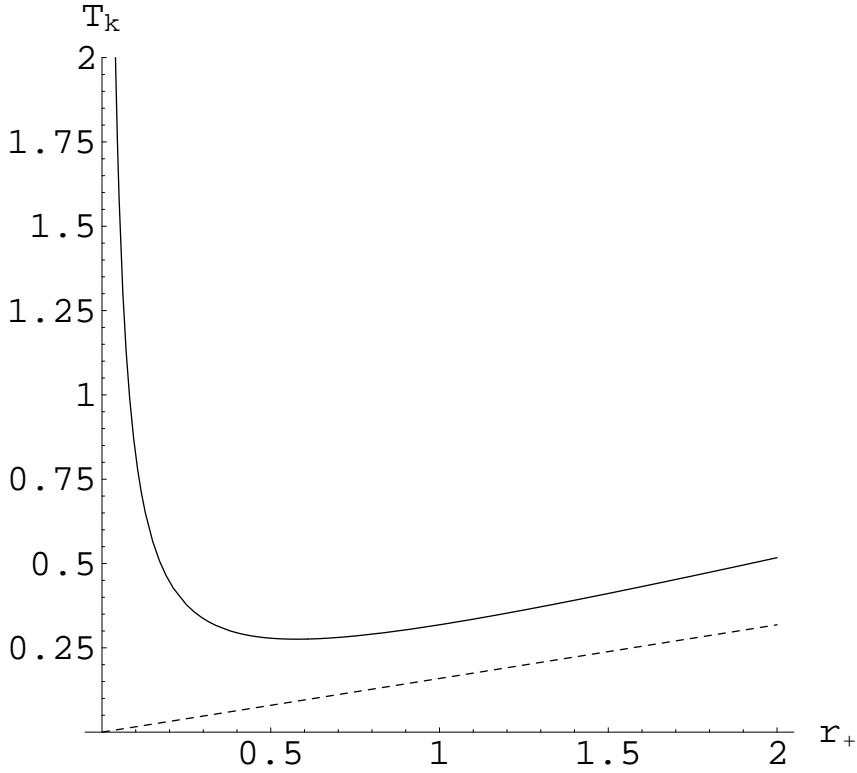,width=0.7\textwidth} \caption{Plot of the
Hawking temperature as a function of the horizon radius $r_+$ with
$l=1$. For $d-2k\not=1$, we have the similar temperature behavior
as in the case of the Schwarzschild-AdS black holes (solid line:
$d=4,k=1$). In this case we have a minimum temperature of
$T^{SADS}_{1}=T_c=\sqrt{3}/2 \pi$ at $r_+=r_c=1/\sqrt{3}=0.58$. On
the other hand, for $d-2k=1$, we have the similar temperature
behavior as in the case of the Chern-Simens black holes (dotted
line: $d=3,k=1$). Its absolute minimum is located  at
$T^{CS}_{1}=T_c=0(r_+=r_c=0)$, which corresponds to the MBTZ
case.} \label{fig1}
\end{figure}

\begin{figure}[t!]
\epsfig{file=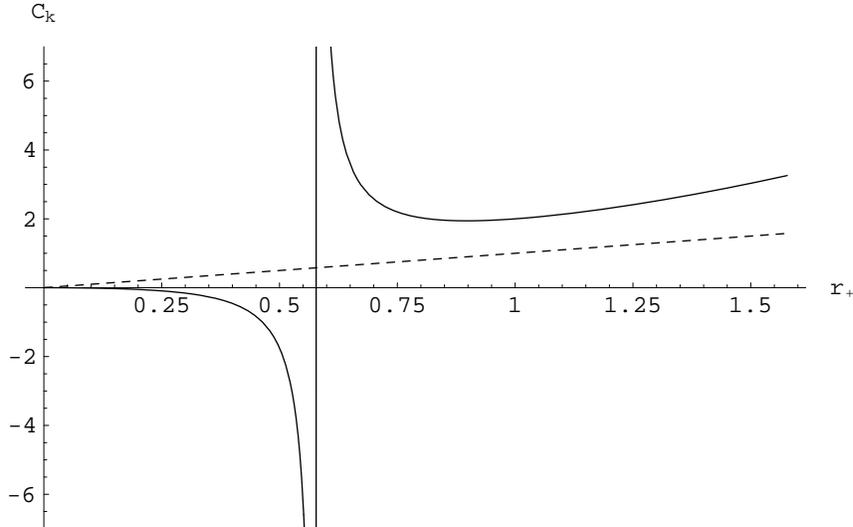,width=0.7\textwidth} \caption{Plot of the
heat capacity as a function of $r_+$ with $l=1$. For
$d-2k\not=1$(solid line: $d=4,k=1$), $C_{1}^{SADS}$ has a pole at
$r_+=r_c=0.58$. This shows that the phase transition at
$T_1=T_c(r_+=r_c)$ is first-order.  On the other hand, for
$d-2k=1$(dotted line: $d=3,k=1$), the heat capacity is a
monotonically continuous increasing function of $r_+$. We have a
zero heat capacity $C_{1}^{CS}=0$ at $r_+=r_c=0$ which corresponds
to the MBTZ case. It shows that the phase transition between NBTZ
with $M \not=0$ and MBTZ with $M=0$ is second-order.} \label{fig2}
\end{figure}

In the case of $d-2k \not=1$, one finds the $d$-dimensional
Schwarzschild-AdS black holes

\beq \label{5eq3}  ds^2_{SADS}=
-\Big[1+\fr{r^2}{l^2}-\Big(\fr{2G_kM}{r^{d-2k-1}}\Big)^{1/k}\Big]dt^2
+\fr{dr^2}{1+\fr{r^2}{l^2}-\Big(\fr{2G_kM}{r^{d-2k-1}}\Big)^{1/k}}+r^2d\Omega_{d-2}
 \eeq
which possesses a continuous mass spectrum from $M(r_+)=
\fr{r_+^{d-2k-1}}{2G_k}\Big(1+\fr{r_+^2}{l^2}\Big)^k$ to the
thermal AdS  with $M(0)=0$:
$ds^2_{ADS}=-(1+r^2/l^2)dt^2+(1+r^2/l^2)^{-1}dr^2+r^2d\Omega_{d-2}^2$.
Its Hawking temperature and  heat capacity are given by\cite{CMN}
\beq \label{6eq3} T_k^{SADS}=\fr{1}{4\pi k}
\Big[\fr{(d-1)r_+}{l^2}+\fr{d-2k-1}{r_+}\Big],~~C_k^{SADS}=\fr{2
\pi k
r_+^{d-2k}}{G_k}\Big(\fr{r_+^2+r_c^2}{r_+^2-r_c^2}\Big)\Big(1+\fr{r_+^2}{l^2}\Big)^{k-1}\eeq
with the Boltzmann constant $k_B=1$ and the minimum length
$r_c=l\sqrt{(d-2k-1)/(d-1)}$. As is shown in Fig. \ref{fig1}, the
Hawking temperature diverges at $r_+=0$. It has the minimum value
of the critical temperature $T_c=(d-1)r_c/2\pi kl^2$ at $r_+=r_c$,
and grows linearly for large $r_+$. Actually the Hawking
temperature for $k\not=1$ and $d\ge 5$ has the similar behavior as
in the $d=4$ Schwarzschild-AdS black hole. As is shown Fig.
\ref{fig2}, the heat capacity has an unbounded discontinuity at
$r_+=r_c$, signaling a phase transition from negative heat
capacity to positive one. Roughly speaking, thermal behavior split
into two branches. For $r_+<r_c$, the heat capacity is negative
and thus the black hole state cannot be in equilibrium with the
heat bath at temperature $T_0$. On the other hand, for $r_+>r_c$,
the heat capacity is positive and thus the black hole state can be
in equilibrium with the heat bath.   If $r_+$ exceeds an unstable
equilibrium position ($r_u$) in the order of $r_u<r_c<r_s$ where
$r_s$ is a locally stable position, the Schwarzschild-AdS black
hole can reach equilibrium with a heat bath of temperature
$T_0>T_c$. Here $T_0$ corresponds to two equilibrium states of
radii $r_u$ and $r_s$\cite{CTZ}. Explicitly, if $r_+>r_u$, the
black hole evolves towards an equilibrium configuration at
$r_+=r_s$.  If either $T_0<T_c$ or $r_+<r_u$, the black hole
continues to evaporate until its final stage. That is, it is
unstable to decay into the thermal AdS space. This is a typical
picture for the first-order Hawking-Page transition in even
dimensions. It shows that the phase transition is sensitive to the
initial black hole state $r_+$ and the temperature $T_0$ of the
heat bath.
 The higher order curvature-correction with $k\not=1$
to the gravity  never alters this feature of phase transition. In
the case of $k=1$, one finds the area-law of the entropy:
$S^{SADS}_{1}=2\pi r_+^{d-2}/(d-2)G_{1}=S_{BH}$. The entropy  for
$k=2$ case takes the form  \beq \label{7eq3} S^{SADS}_{2} =\fr{4
\pi}{G_k}\Big[ \fr{r_+^{d-2}}{(d-2)l^2}+\fr{r_+^{d-4}}{d-4} \Big],
\eeq where one cannot find the area-law behavior because of the
presence of the last term.

 The other case is for odd dimensions with
$d-2k=1(k=n-1)$, called the Chern-Simons black holes as \beq
\label{8eq3} ds^2_{CS}=
-\Big[1+\fr{r^2}{l^2}-\Big(2G_kM+1\Big)^{1/k}\Big]dt^2
+\fr{dr^2}{1+\fr{r^2}{l^2}-\Big(2G_kM+1\Big)^{1/k}}+r^2d\Omega_{d-2}
 \eeq
which possesses a continuous mass spectrum from $M(r_+)=
\fr{r_+^{d-2k-1}}{2G_k}\Big(1+\fr{r_+^2}{l^2}\Big)^k-\fr{1}{2G_k}$
to the massless AdS black holes ($M=0$) with different topology:
$ds^2_{MADS}=-(r^2/l^2)dt^2+(r^2/l^2)^{-1}dr^2+r^2d\Omega_{d-2}^2$.
The  temperature and heat capacity  are  given by, respectively,
\beq\label{9eq3} T_k^{CS}=\fr{1}{4\pi k}
\fr{(d-1)r_+}{l^2},~~C_k^{CS}=\fr{2 \pi k
r_+}{G_k}\Big(1+\fr{r_+^2}{l^2}\Big)^{k-1}.\eeq In this case the
Hawking temperature and heat capacity are not divergent at all.
These both are monotonically increasing positive  functions of
$r_+$ (see Fig. 1 and 2). The minimum temperature is given by
$T_k^{CS}(r_+=0)=T_c=0$ for the MADS case. Roughly speaking, the
presence of the negative cosmological constant $\Lambda$ makes it
possible for the Schwarzschild-AdS black holes  to reach the
thermal equilibrium under the condition of   $T_0>T_c$ and
$r_+>r_u$. In the case of the Chern-Simons black holes, the heat
capacity is always positive and therefore the equilibrium
configuration is always reached, regardless of the initial black
hole state $r_+$ and the temperature $T_0$ of the heat reservoir.
The entropy is defined from the Euclidean path integral as \beq
\label{10eq3} S_k^{CS}=\fr{2 \pi k}{G_k} \int^{r_+}_0
\Big(1+\fr{r^2}{l^2}\Big)^{k-1} dr. \eeq This  is also a
monotonically increasing functions of $r_+$. For $k=2$ case, we
have \beq \label{11eq3} S_2^{CS}=\fr{4 \pi r_+}{G_k}
\Big(1+\fr{r_+^2}{3l^2}\Big). \eeq

  In case of the Schwarzschild-AdS black
holes, there exists a first-order phase transition at
$T^{SADS}_k=T_c (r_+=r_c)$. The heat capacity $C^{SADS}_k$ is
positive for $r_+>r_c$, while it is negative for $r_+<r_c$. Also
$C^{SADS}_k$ has a simple pole at $r_+=r_c$.  This confirms from
the analysis of the free energy ($F=M-T_HS$), although its exact
transition point of $r_+=r_c=l$  for $k=1$  is different from
$r_+=r_c=l/\sqrt{3}$ derived by the heat capacity. We note  a sign
change between \beq \label{12eq3} F(r_+ \to 0) \sim
r_+^{d-2k-1}/2(d-2k)G_k ~{\rm and}~ F(r_+ \to \infty) \sim
-r_+^{d-1}/2(d-2)G_kl^{2k} \eeq which shows  that
 a small black hole with $r_+<r_c$ is unstable to decay into
thermal AdS space, while  a large black hole with $r_+>r_c$ is
stable.

However, we always have a positive heat capacity of $C^{CS}_k>0$
for the Chern-Simons black holes. Since the Chern-Simons black
holes are exceptional class, there is no phase transition at all
even though $F$ has a change in sign as is shown in
Eq.(\ref{12eq3}).  This disagreement comes from a difference
between the canonical ensemble approach with fixed temperature and
the  free energy approach with  variable temperature. Actually,
there is  no critical temperature $T_c$ which gives us a
first-order phase transition in the Chern-Simons system. They are
genuine gauge theories whose
 supersymmetric extension is known.
 These black holes can reach thermal equilibrium with a heat bath
 at any temperature. {\it The positivity of heat capacity guarantees
 their stability under thermal fluctuations.} If the heat capacity is negative,
 there is no stable thermal fluctuations\cite{ACB}. Accordingly, a small
 Chern-Simons black hole with $r_+<l$ is stable against decay by the Hawking
 radiation. In three dimensions, this is known to be  the non-rotating BTZ black
 hole (NBTZ). One
recovers
 thermodynamic quantities for the NBTZ with $G_1=1/2$ \beq
\label{13eq3} T_1^{CS} \to T_H=\fr{r_+}{2 \pi l^2},~~C_1^{CS}\to
C_J=4 \pi r_+,~~S_1^{CS} \to S_{BH}=4 \pi r_+. \eeq  The NBTZ
spectrum  has a mass gap separating it from thermal AdS space. We
are interested in constructing the Gott time machine to study the
formation of the non-rotating BTZ black hole by a two-body
collision process\cite{BB}. The NBTZ is defined by a hyperbolic
isometry. We recall that the mass of NBTZ is $M>0$, while the
point particle mass $m$ is related to $M$ by $m=2(1-\sqrt{-M})$.
Then the point particle spectrum is given by $-1<M<0$ which
belongs to a branch  of  conical deficits. Here $M=0$ corresponds
to MBTZ and $M=-1$ is TADS.   Let us introduce  the static black
hole with the same left and right generators. We choose
$\rho_L=\rho_R=-T^G=\rho$ because isometries of the  NBTZ are
subject to  identifications with $(\rho_L,\rho_R) \sim
(-\rho_L,-\rho_R)$. If the Gott condition is satisfied, $T^G$ is a
hyperbolic generator and thus the Gott time machine results in
formation of the NBTZ. The mass  is given by an input  parameter
$p$ which depends on the initial data \beq \label{1eq4} \fr{{\rm
Tr}\rho}{2} =\cosh \Big[ \pi \sqrt{M}\Big] \equiv p,~~ p\ge
p_*=1\eeq which means that an order parameter for formation of
NBTZ is the trace of the generator. This takes a critical value at
the threshold for the black hole formation ($p=p_*$). In other
words, $p=p_*$ corresponds to the vacuum of the black hole (MBTZ),
where the Gott generator becomes parabolic. From Eq.(\ref{1eq4}),
one finds an exact  formula for the formation of NBTZ in terms of
the parameter $p$ as \beq \label{2eq4} \pi \sqrt{M}=\cosh^{-1}
(p)=\ln\Big[p+\sqrt{p^2-1}\Big].\eeq From the above,  one
determines the Choptuik scaling for formation of the NBTZ \beq
\label{3eq4} \sqrt{M} =\fr{r_+}{l}
=\fr{\sqrt{2}}{\pi}(p-p_*)^{1/2},\eeq where implies that the
Choptuik scaling exponent  is given by $\gamma=1/2$.

If the Gott condition is not satisfied, one has an effective
particle spacetime with an elliptic generator. In case of
$-1<M<0$, a conical  deficit angle $\alpha$ is introduced for an
order parameter to describe the phase transition. Defining $p$ as
\beq \label{4eq4} \fr{{\rm Tr}\rho}{2} =\cos\Big[ \pi
\sqrt{-M}\Big] \equiv p,~~ p< p_*=1\eeq one has similarly the
Choptuik scaling exponent with $\gamma=1/2$ as \beq \label{5eq4}
\alpha =2 \pi -2\sqrt{2}(p_*-p)^{1/2}.\eeq In case of MBTZ with
$M=0$, we have $p=p_*$ and $\alpha=2\pi$. For TADS with $M=-1$, we
have $\alpha=2\pi-4>0$.

 \sect{No Hawking-Page phase transition in three dimensions}

In $d\ge 4$ case, Hawking-Page phase transition occurs between
Schwarzschild-AdS black hole and thermal AdS space. This is
possible because the presence of a negative cosmological constant
is sufficient to render the canonical ensemble well defined for
all Schwarzschild-AdS black holes. However, for  Chern-Simons
black holes (NBTZ case), the situation is quite different from
those of Schwarzschild-AdS black holes. For our purpose, we
express the free energy, energy and heat capacity in terms of the
Hawking temperature as \beq \label{1eq5}
F_{NBTZ}=-E_{NBTZ}=-\Big(2\pi l \Big)^2
T_H^2,~~C_{NBTZ}=S_{NBTZ}=2\Big(2\pi l\Big)^2 T_H. \eeq It is
obvious that the NBTZ with $T_H=0$ leads to those for MBTZ case as
\beq \label{2eq5} F_{MBTZ}=E_{MBTZ}=C_{MBTZ}=S_{MBTZ}=0. \eeq On
the other hand, thermodynamic quantities for thermal AdS space are
given by \beq \label{3eq5}
F_{TADS}=E_{TADS}=-1,~~C_{TADS}=S_{TADS}=0 \eeq with $T_H=0$.

\begin{figure}[t!]
\epsfig{file=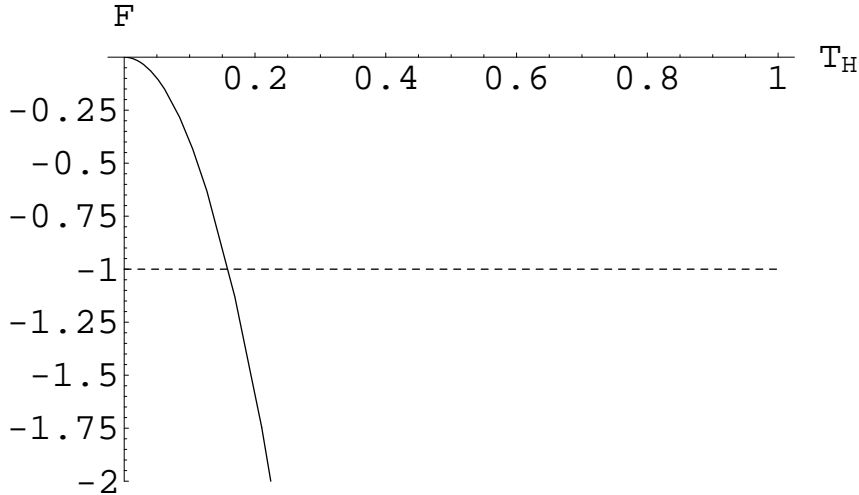,width=0.7\textwidth} \caption{Plot of the
free energy as function of the Hawking temperature $T_H$ with
$l=1$. The solid line represents the free energy behavior for the
NBTZ case, while the dotted line denotes the constant negative
free energy for the TADS case. Their values coincide with each
other at the assumed critical temperature $T_H=T_c=1/2\pi$.}
\label{fig3}
\end{figure}

\begin{figure}[t!]
\epsfig{file=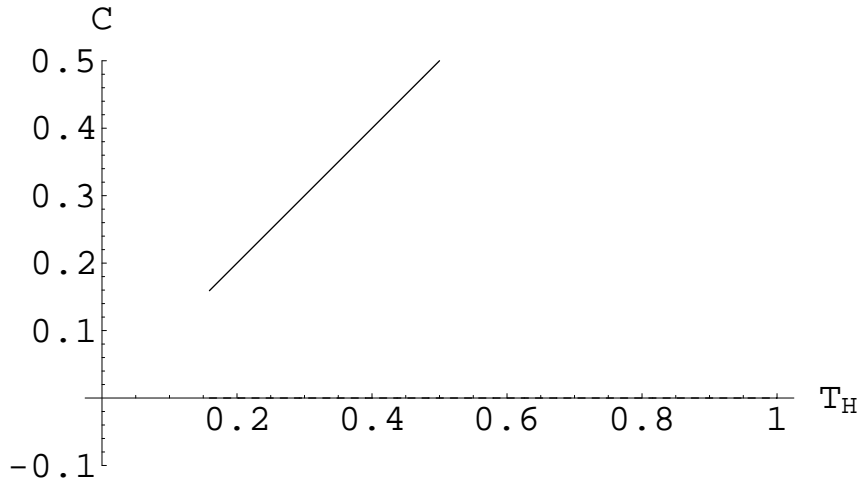,width=0.7\textwidth} \caption{Plot of the
heat capacity  as a function of the Hawking temperature with
$l=1$. Below the assumed critical temperature $T_H=T_c=1/2 \pi$,
the heat capacity for the TADS is zero.  The heat capacity for
NBTZ increases linearly for $T_H>T_c$. Here we have a finite jump
of the heat capacity at $T_H=T_c$. However, even if the assumed
transition occurs really, it does not imply the presence of the
first-order Hawking-Page transition.} \label{fig4}
\end{figure}
At this stage, we introduce  the assumed  picture of the
first-order Hawking-Page transition in three dimensions. A
first-order phase transition may occur at $T_H=T_c=1/2\pi l$
between NBTZ and TADS\cite{KS}. As is shown in Fig. \ref{fig3},
for $T_H<T_c$, the free energy of TADS is lower than that of NBTZ
so that  NBTZ is less probable than TADS. For $T_H>T_c$, the NBTZ
is more probable than TADS.  We raise the temperature of system
gradually from the lower one.  Then  TADS dominates as long as
$T_H<T_c$. When $T_H=T_c$, the TADS begins to be transformed into
the NBTZ configuration. Also  there exists discontinuity for
energy and heat capacity between NBTZ and TADS. However, in three
dimensions, one has  a mass gap between
 MBTZ  with $M=0$ and TADS with $M=-1$.  A conical deficit
interpreted as a point mass source appears between these. As was
shown in the previous section, the branch of $-1<M<0$ is
 totally different from the branch of NBTZ with $M>0$. A
 second-order phase transition might  be taken from MBTZ with $M=0$ to
NBTZ with $M>0$ because the  energy and heat capacity (entropy)
are continuous\cite{CLZ}. This calculation is based on the
semiclassical approximation for the black hole thermodynamics.

Alternatively, if one includes quantum fluctuations, there exits a
possibility that the MBTZ is not the end of the Hawking
evaporation and the end might be the TADS\cite{LO}. Then we may
introduce the above assumed transition between NBTZ and TADS. It
may correspond to a counterpart to the first-order Hawking-Page
transition between Schwarzschild-AdS black hole with $M>0$ and
thermal AdS space with $M=0$. However, this phase transition turns
out to be second-order because the heat capacity has not a pole
but a finite jump at the assumed critical temperature $T_H=T_c$.

Consequently,  the  transitions between NBTZ and MBTZ (TADS) might
be  possible to occur in three dimensions but these do not belong
to the first-order Hawking-Page transition.

\section*{Acknowledgement}
The author thanks to Gungwon Kang and Jungjai Lee for helpful
discussions at the early stage of this work. This work was
supported by the Korea Research Foundation
Grant(KRF-2005-013-C00018).

\end{document}